\documentclass[aps,prl,showpacs,twocolumn,amsmath,amssymb,superscriptaddress]{revtex4}

\usepackage{graphicx}
\usepackage{multirow}
\usepackage{dcolumn}
\usepackage{bm}

\begin{document}

\title{Suppression of Grain Boundaries in Graphene Growth on Superstructured Mn-Cu(111) Surface}

\author{Wei Chen}
   \affiliation{Department of Physics and Astronomy, University of Tennessee, Knoxville, TN 37996, USA}
   \affiliation{ICQD/HFNL, University of Science and Technology of China, Hefei, Anhui, 230026, China}

\author{Hua Chen}
   \affiliation{Department of Physics and Astronomy, University of Tennessee, Knoxville, TN 37996, USA}
   \affiliation{Department of Physics, University of Texas, Austin, TX 78712, USA}

\author{Haiping Lan}
   \affiliation{ICQD/HFNL, University of Science and Technology of China, Hefei, Anhui, 230026, China}

\author{Ping Cui}
   \affiliation{ICQD/HFNL, University of Science and Technology of China, Hefei, Anhui, 230026, China}

\author{Tim P. Schulze}
   \affiliation{Department of Mathematics, University of Tennessee, Knoxville, TN 37996, USA}

\author{Wenguang Zhu}
   \affiliation{Department of Physics and Astronomy, University of Tennessee, Knoxville, TN 37996, USA}
   \affiliation{Materials Science and Technology Division, Oak Ridge National Laboratory, Oak Ridge, TN 37831, USA}

\author{Zhenyu Zhang}
   \affiliation{ICQD/HFNL, University of Science and Technology of China, Hefei, Anhui, 230026, China}
   \affiliation{School of Engineering and Applied Sciences, Harvard University, Cambridge, MA 02138, USA}
\date{\today}

\begin{abstract}
As undesirable defects, grain boundaries (GBs) are widespread in epitaxial graphene using existing growth methods on metal substrates. Employing density functional theory calculations, we first identify that the misorientations of carbon islands nucleated on a Cu(111) surface lead to the formation of GBs as the islands coalesce. We then propose a two-step kinetic pathway to effectively suppress the formation of GBs. In the first step, large aromatic hydrocarbon molecules are deposited onto a $\sqrt{3}\times\sqrt{3}$ superstructured Cu-Mn alloyed surface to seed the initial carbon clusters of a single orientation; in the second step, the seeded islands are enlarged through normal chemical vapor deposition of methane to form a complete graphene sheet. The present approach promises to overcome a standing obstacle in large scale single-crystal graphene fabrication.
\end{abstract}

\pacs{81.05.ue, 81.15.-z, 68.43.-h, 68.35.bd}

\maketitle

Graphene is a one-atom-thick flat sheet of carbon atoms packed into a honeycomb structure. Because of its superb mechanical, electronic, optical, and thermal properties~\cite{geim_natmater_2007,geim_science_2009}, graphene has limitless potential for future device applications. To fully realize the functionality of graphene, it is highly desirable to fabricate large-scale monolayer graphene with no or minimal structural defects. Among the different fabrication avenues being explored, epitaxial growth on transition metal substrates using hydrocarbon or other carbon sources stands out as a highly appealing approach~\cite{kim_nature_2009,coraux_newjphys_2009,sutter_natmater_2008,sutter_prb_2009,ago_acsnano_2010,kwon_nanolett_2009,roth_surfsci_2011,oznuluer_apl_2011,li_science_2009}, especially on Cu surfaces~\cite{li_science_2009}. Cu has the merit of low carbon solubility, which leads to a self-limiting growth process confined to its surface~\cite{li_nanolett_2009}, and diverse carbon sources can be used to grow graphene on Cu substrates~\cite{gao_nanolett_2010,sun_nature_2010,li_acsnano_2011}. The relatively weak carbon-copper interaction compared to carbon-carbon interaction enables fast diffusion of carbon atoms and efficient nucleation of carbon islands across the whole surface~\cite{chen_prl_2010}, indicating the feasibility of mass production of epitaxial graphene. Indeed, it has been reported recently that the single-crystal domains of monolayer graphene grown on Cu can reach the dimensions of 0.5mm on a side~\cite{li_jacs_2011}. Furthermore, transfer of epitaxial graphene to other substrates can be readily achieved via chemical etching~\cite{li_nanolett_transfer_2009,yu_natmater_2011}.

However, one standing obstacle facing the community of epitaxial graphene on Cu is the prevalence of grain boundaries (GBs) undesirably introduced during growth~\cite{yu_natmater_2011,ajayan_natmater_2011,yazyev_prb_2010,yazyev_natmater_2010,grantab_science_2010,huang_nature_2011,kim_acsnano_2011,an_acsnano_2011}. A grain boundary refers to the junction region of two crystalline grains with different orientations. The detailed atomic structures of the GBs in epitaxial graphene have been investigated extensively~\cite{yu_natmater_2011,yazyev_prb_2010,yazyev_natmater_2010,grantab_science_2010,huang_nature_2011,kim_acsnano_2011,an_acsnano_2011}, and their presence has been shown to severely degrade the electronic, transport, and mechanical properties of graphene~\cite{yu_natmater_2011,huang_nature_2011}. Experimental efforts have also been made to suppress their creation during growth~\cite{li_jacs_2011,yu_natmater_2011,li_nanolett_2010}, but so far with limited success, partly because the underlying formation mechanism of the GBs is still unclear. Existing experimental observations suggest that the GBs can form in the initial nucleation stage when several graphene grains emanate from one nucleation site~\cite{huang_nature_2011}; alternatively, they can be formed in the later growth stage when different graphene grains with relative misorientations coalesce~\cite{gao_nanolett_2010,yu_natmater_2011,an_acsnano_2011}.

In this letter, we first demonstrate that, because of the inherently weak C-Cu interaction, orientational disorders of carbon islands on Cu(111) will be abundant in the early stages of nucleation and growth. Such disorders cannot heal themselves with the enlargement of the islands, leading to the prevalence of graphene GBs upon island coalescence. Based on this understanding, we propose to use a functionalized Cu(111) surface to lift the energy degeneracy in the early stages of nucleation and growth, thereby suppressing orientational disorders of the islands and the subsequent GBs. Our proposed kinetic pathway invokes the steps of ``seed and grow''~\cite{rosenfeld_prl_1993}. In the seeding step, carbon clusters are initiated by depositing coronene~\cite{fetzer_2000} on a ($\sqrt{3}\times\sqrt{3}$) R30$^{\circ}$ Mn-Cu(111) alloyed surface~\cite{schneider_ass_1999,bihlmayer_prb_2000}, which effectively helps the islands to select predominantly only one orientation on the superstructurally alloyed surface. In the growing step, larger, monolayer graphene is formed by conventional chemical vapor deposition (CVD).

Our density functional theory (DFT) calculations are carried out using the Vienna \textit{ab initio} simulation package (VASP)~\cite{kresse_prb_1996} with PAW potentials~\cite{blochl_prb_1994,kresse_prb_1999} and the generalized gradient approximation (PBE-GGA)~\cite{perdew_prl_1996} for the exchange-correlation functional. The lattice constant of Cu is obtained via structural optimization. The generic Cu(111) surface is modeled by a slab of 5 atomic layers, and the Mn-Cu(111) surface is realized by substituting Cu atoms by Mn atoms at ordered positions in the first layer. The vacuum layers are more than 13 {\AA} thick to ensure decoupling between neighboring slabs. During relaxation, atoms in the lower 2 atomic layers are fixed in their respective bulk positions, and all the other atoms are allowed to relax until the forces on them are smaller than 0.01 eV/{\AA}. A 2$\times$2$\times$1 $k$-point mesh is used for the 6$\times$6 surface unit cell and 3$\times$3$\times$1 for the 4$\times$4 surface unit cell~\cite{methfessel_prb_1989}. The calculations with Mn atoms are spin-polarized. We consider the ferromagnetic configuration of the Mn-Cu(111) surface, because of the triangular arrangement of the Mn atoms. The binding energies are calculated as  $\Delta E = E_\mathrm{adsorbate} + E_\mathrm{substrate} - E_\mathrm{adsorbate+substrate}$.

\begin{figure}
 \begin{center}
\includegraphics[width=2.5 in]{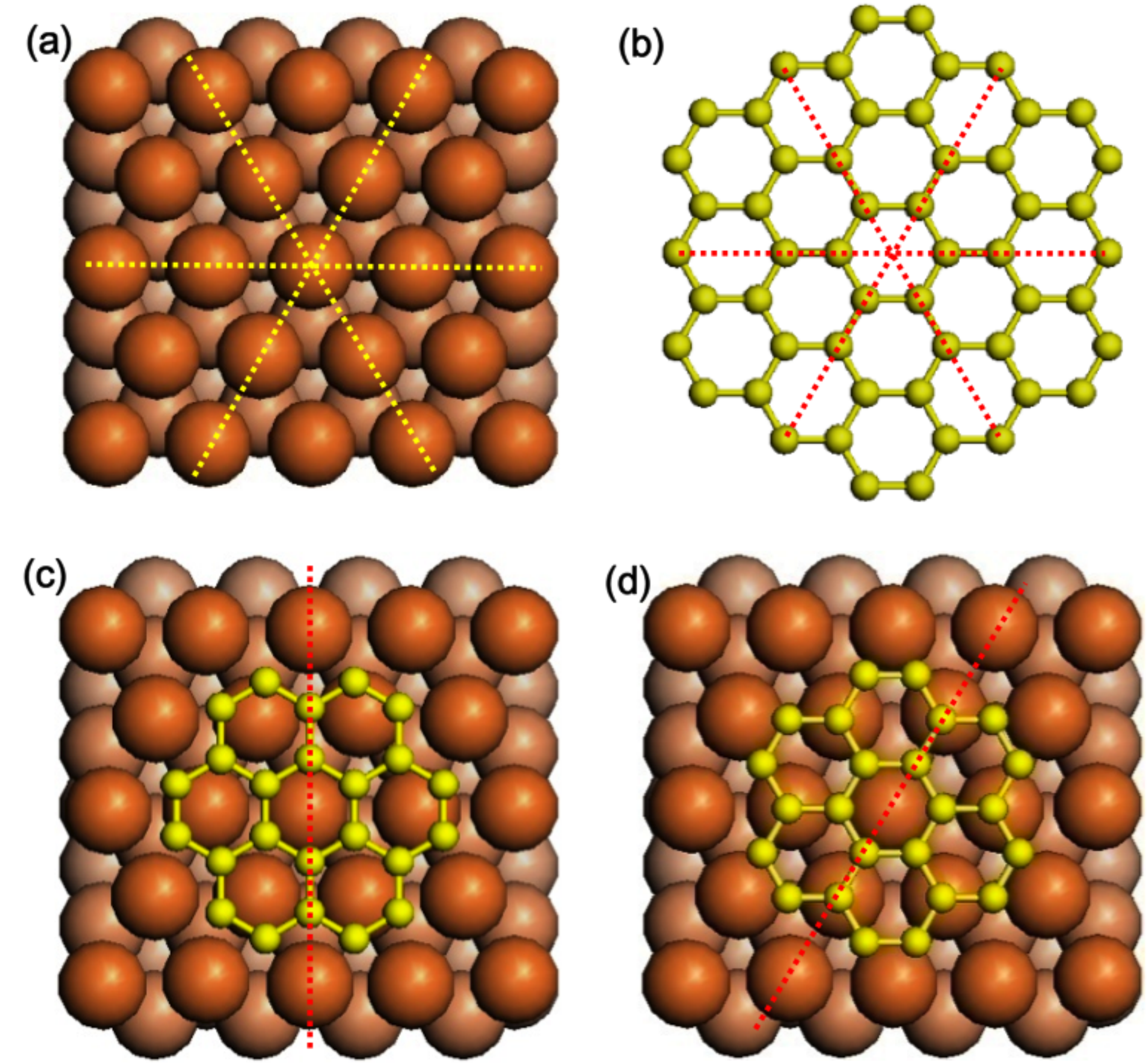}
 \end{center}
 \caption{(color online). (a) and (b) Structural illustrations of the Cu (111) substrate and graphene, where the yellow and red dashed lines show their respective high-symmetry axes. (c) and (d) Illustrations of two geometries where a 7CR carbon cluster is at a HSO on the Cu(111) surface. In (c), the edge C atoms reside at the 3-fold hollow sites; in (d), the edge C atoms are at the bridge sites between two surface Cu atoms.}
 \label{figure1}
\end{figure}

Crystalline Cu has a face-centered cubic (fcc) structure, and its (111) surface exhibits a hexagonal packing of surface atoms. As shown in Figs. 1 (a) and (b), both the Cu(111) surface and graphene have atomic arrangements with six-fold symmetry. Therefore, if carbon clusters nucleated at different sites are all oriented at the \textit{same} high-symmetry orientation (HSO) of the Cu(111) surface (e.g., Fig. 1 (c) \textit{or} (d)), their structural coherence will be ensured by the Cu substrate and there will be no GBs when they merge. However, when a simple six-fold symmetric carbon cluster composed of seven 6-carbon rings (7CR) is placed on the Cu(111) surface, our detailed DFT calculations reveal that the energetically most stable geometry deviates from the HSO of Fig. 1 (c) by 11$^{\circ}$ (Fig. 2 (b)). The carbon cluster also has a dome-like structure (Fig. 2 (a)), with the central C atoms $\sim$2.30 {\AA} from the Cu surface. Therefore, the cluster remains strongly bonded to the substrate only at the periphery while the interaction between the central C atoms and the substrate is rather weak, similar to the domed structure on Ir(111)~\cite{lacovig_prl_2009}. Each of the 12 edge C atoms has two C neighbors, and prefers to reside at the bridge sites between two surface Cu atoms, because these edge atoms are closer to $sp^3$ hybridization than $sp^2$, thereby providing the driving force for the rotation of the island away from the HSO.

\begin{figure}
 \begin{center}
\includegraphics[width=3.0 in]{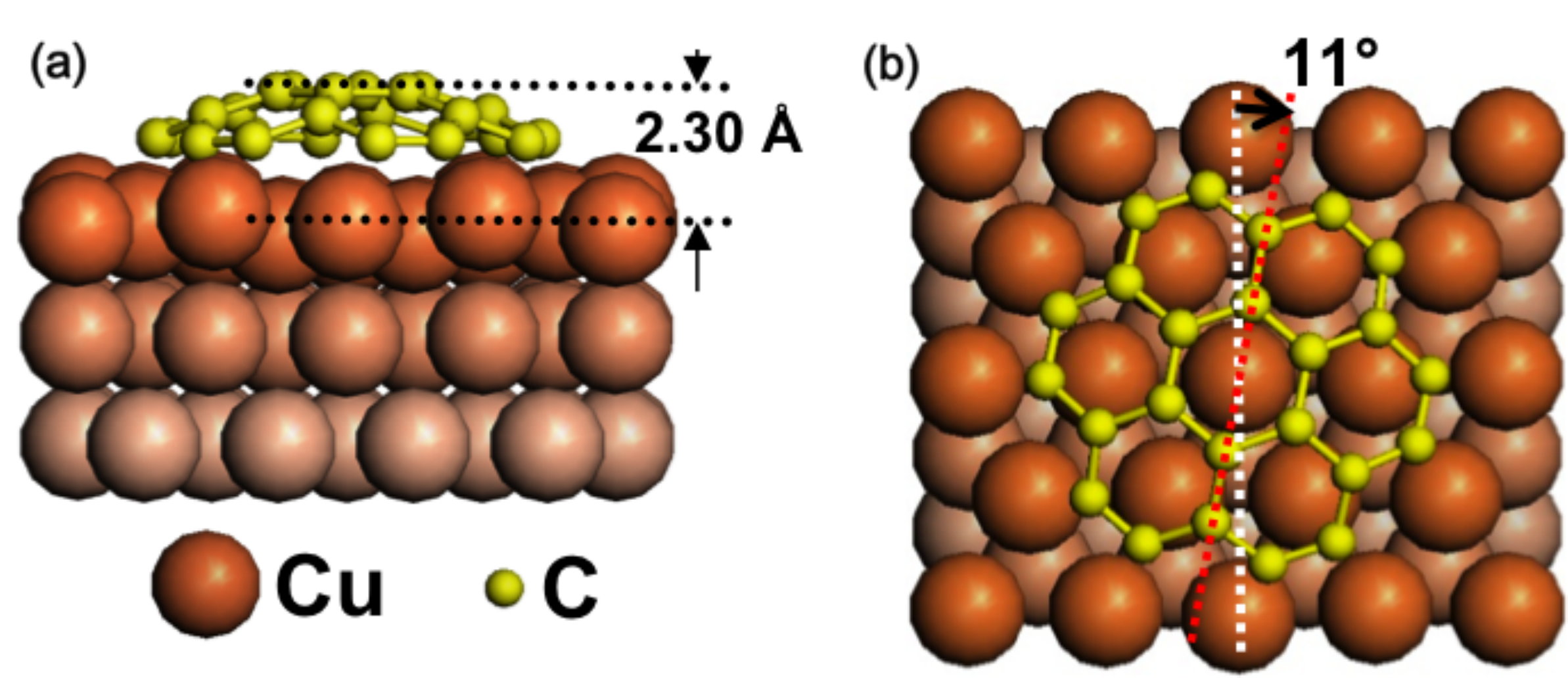}
 \end{center}
 \caption{(color online). Side and top view of a 7CR on the Cu(111) surface, illustrating the domed nature (a) and the rotated nature (b) from the HSO of Fig. 1 (c), respectively.}
 \label{figure2}
\end{figure}

Now we go from the early stages of island nucleation and growth to island enlargement and coalescence. Since a small cluster such as the one shown in Fig. 2 is not oriented at the HSO, there will be a degenerate mirror geometry with respect to the symmetry axis of Cu(111), indicating that islands with relative misorientations can coexist. As a cluster grows larger, more edge C atoms will be involved in determining its preferred orientation by adjusting their bonds with the underlying Cu atoms. Therefore, there will be more nearly degenerate orientations, thus broadening the orientational disorder of the carbon clusters. When a cluster has grown large enough such that the edge atoms contribute only minimally to the total binding energy, the cluster is either still in an energetically stable orientation different from the HSO, or is too large to adjust its orientation to an energetically more favorable HSO. When two such clusters with a relative misorientation coalesce, a larger island containing a GB is formed, with a characteristic angle defined by the initial misorientations of the merging clusters and the local structural adjustment within the boundary. This scenario is qualitatively consistent with existing experimental observations, and the detailed distribution of the GB angles may also depend on the specific growth conditions~\cite{huang_nature_2011,kim_acsnano_2011,an_acsnano_2011}.

Next we search for ways to lift the energy degeneracy in island nucleation and growth via Cu(111) surface modification. Fig. 3 (a) illustrates a ($\sqrt{3}\times\sqrt{3}$) R30$^{\circ}$ X-Cu(111) superstructurally alloyed surface, which has transition metal atoms X substituting Cu at ordered positions. The choice of X is guided by the requirements that, (a) C binds more strongly to X than to Cu, so that the nucleated carbon clusters will prefer a HSO in maximizing their interaction with the X atoms; and (b) the alloyed X-X atoms are repulsive. We find that a number of transition metals (X= Ru, Fe, Co, Ni, Mn) have unfilled \textit{d} orbitals and bind more strongly with C, but only Mn atoms are mutually repulsive when alloyed into the top layer of Cu(111)~\cite{chen_prepare}, indicating that Mn could form a superstructured surface alloy with Cu. Indeed, experimentally the formation of a ($\sqrt{3}\times\sqrt{3}$) R30$^{\circ}$ superstructured Mn-Cu(111) surface has been observed~\cite{schneider_ass_1999}, whereas the other metals only form islands or overlayers on Cu surfaces~\cite{chen_ssp_2008}.

\begin{figure}
 \begin{center}
\includegraphics[width=2.5 in]{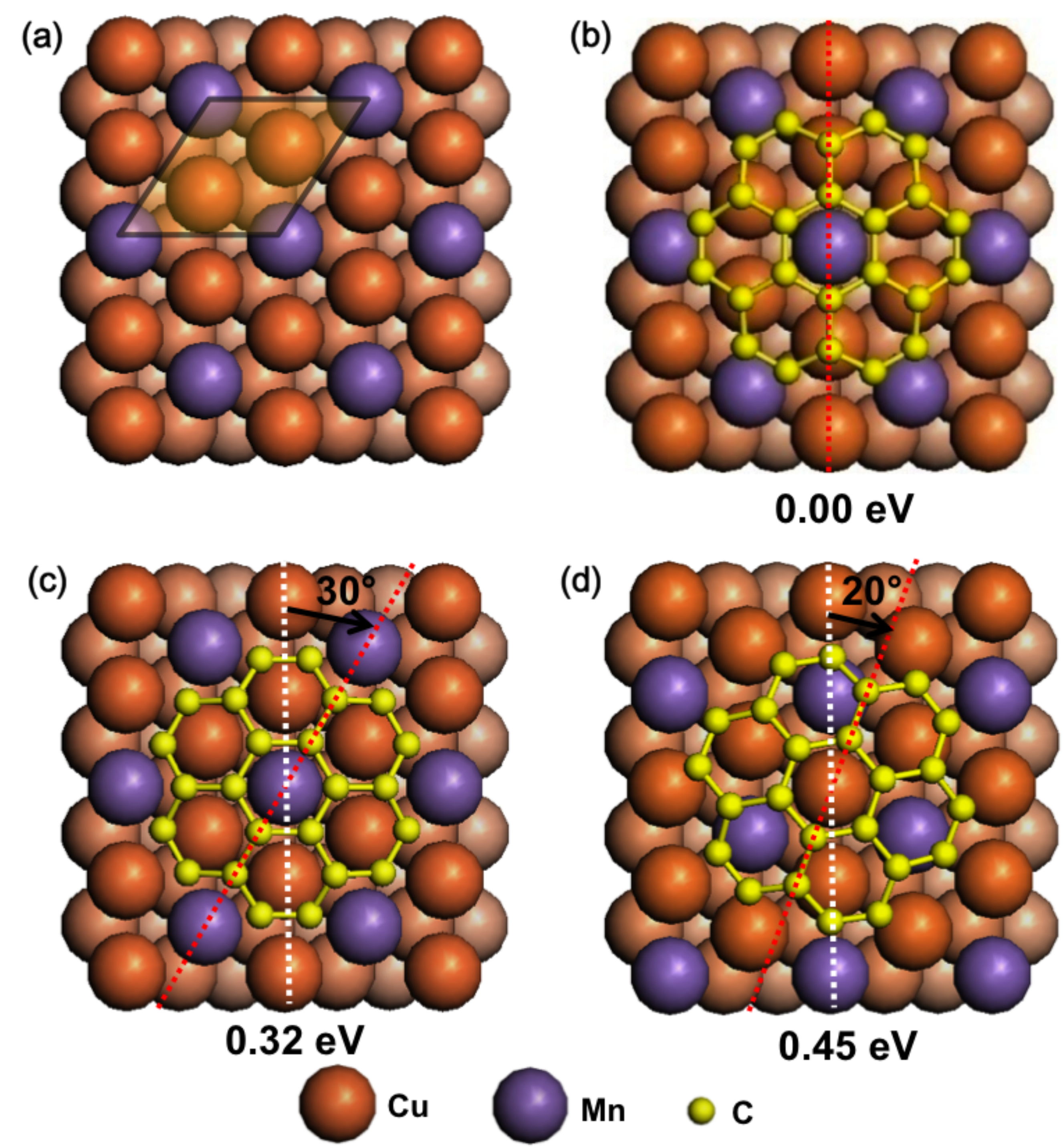}
 \end{center}
 \caption{(color online). (a) Structure of the ($\sqrt{3}\times\sqrt{3}$) R30$^{\circ}$ Mn-Cu(111) superstructurally alloyed surface. (b-d) Three stable geometries of a 7CR adsorbed on the Mn-Cu(111) surface. As indicated by the red dashed lines, (b) and (c) illustrate two HSO configurations, while (d) is rotated from a HSO, and their relative stabilities are indicated by their total energy differences.}
 \label{figure3}
\end{figure}

Still choosing 7CR as the testing baby graphene, we then calculate the total energies of 7CR with different orientations on the ($\sqrt{3}\times\sqrt{3}$) R30$^{\circ}$ Mn-Cu(111) alloyed surface. We find three stable or metastable configurations of a 7CR island through structural optimization, differentiated by placing the center of the 7CR above a Mn or Cu atom, as shown in Fig. 3. Two of them (Figs. 3 (b) and (c)) are at HSO, but only the HSO in Fig. 3 (b) is the most stable, while the energy of the other two configurations is higher by 0.32 eV and 0.45 eV, respectively. Therefore, the Mn atoms alloyed into the Cu(111) surface indeed successfully help to pin the 7CR at the HSO. To see the underlying atomistic reason, we note that in all the three cases, the island has a dome-like geometry similar to that on a pure Cu(111) surface, indicating the predominant interaction with the substrate at its edge. Moreover, in the most stable configuration, the 7CR maximizes its contact with the Mn atoms at the periphery. The calculated binding energy per edge C atom of 7CR is 0.63 eV on Cu(111) and 0.89 eV on Mn-Cu(111).

To take advantage of the superstructural Mn-Cu(111) surface and effectively suppress the possible disorders induced in the initial nucleation process, we propose the use of coronene as a good candidate to seed the initial carbon clusters. As a polycyclic aromatic hydrocarbon, coronene~\cite{fetzer_2000} is just like the 7CR island, but with a hydrogen atom on each edge C atom. We have compared the dehydrogenation process of coronene to that of benzene, which has been used for low temperature graphene growth on Cu~\cite{li_acsnano_2011}. In free space, the C-H bond dissociation energies of benzene and coronene are very close~\cite{barckholtz_jacs_1999}, as verified also by our DFT calculations. Considering the effect of the catalytic substrates, the energy difference between the initial state, where a coronene is adsorbed onto the Mn-Cu alloyed surface, and the final state, where all the edge hydrogen atoms are dissociated in the form of H$_2$, is 1.23 eV per H atom, which is 0.16 eV lower than that of benzene on Cu(111). Therefore, the alloyed surface will be more catalytic in dehydrogenating adsorbed coronene than Cu(111) in dehydrogenating benzene~\cite{li_acsnano_2011}. We therefore propose to use coronene as the first-step carbon source to seed carbon clusters on the patterned surface.

After the deposition and dehydrogenation of coronene on Mn-Cu(111) surface, all the adsorbed 7CR islands will have the same orientation. In particular, when two such 7CR islands coalesce to form a larger graphene cluster, no GB is formed (Fig. 4). In order to achieve a continuous sheet of graphene, we invoke a second step of growth to supply C atoms to fill the openings between the carbon clusters. We notice that, in contrast to Cu(111), on which C adatoms are energetically much more favorable to nucleate than to stay apart~\cite{chen_prl_2010}, here on Mn-Cu(111) the carbon adatoms are less strongly inclined to nucleate. The energy difference between a C dimer and two C monomers on the substrate, calculated as $\Delta E = E_\mathrm{monomer+substrate}\times2 - E_\mathrm{substrate} - E_\mathrm{dimer+substrate}$ is 2.92 eV on Cu(111) and becomes 1.70 eV on Mn-Cu(111). Therefore, the conventional CVD growth using methane or ethylene could be applied here to supply carbon adatoms to diffuse and attach to the nearby coronene-seeded and correctly-oriented carbon islands (Fig. 4), rather than to nucleate new islands, which is similar to the enhanced layer-by-layer growth of Ag on Ag(111) via a two-step kinetic pathway~\cite{rosenfeld_prl_1993}. Eventually, the 7CR-seeded islands will be enlarged and connected to achieve a single-crystal graphene sheet with no or greatly suppressed GBs.

\begin{figure}
 \begin{center}
\includegraphics[width=2.5 in]{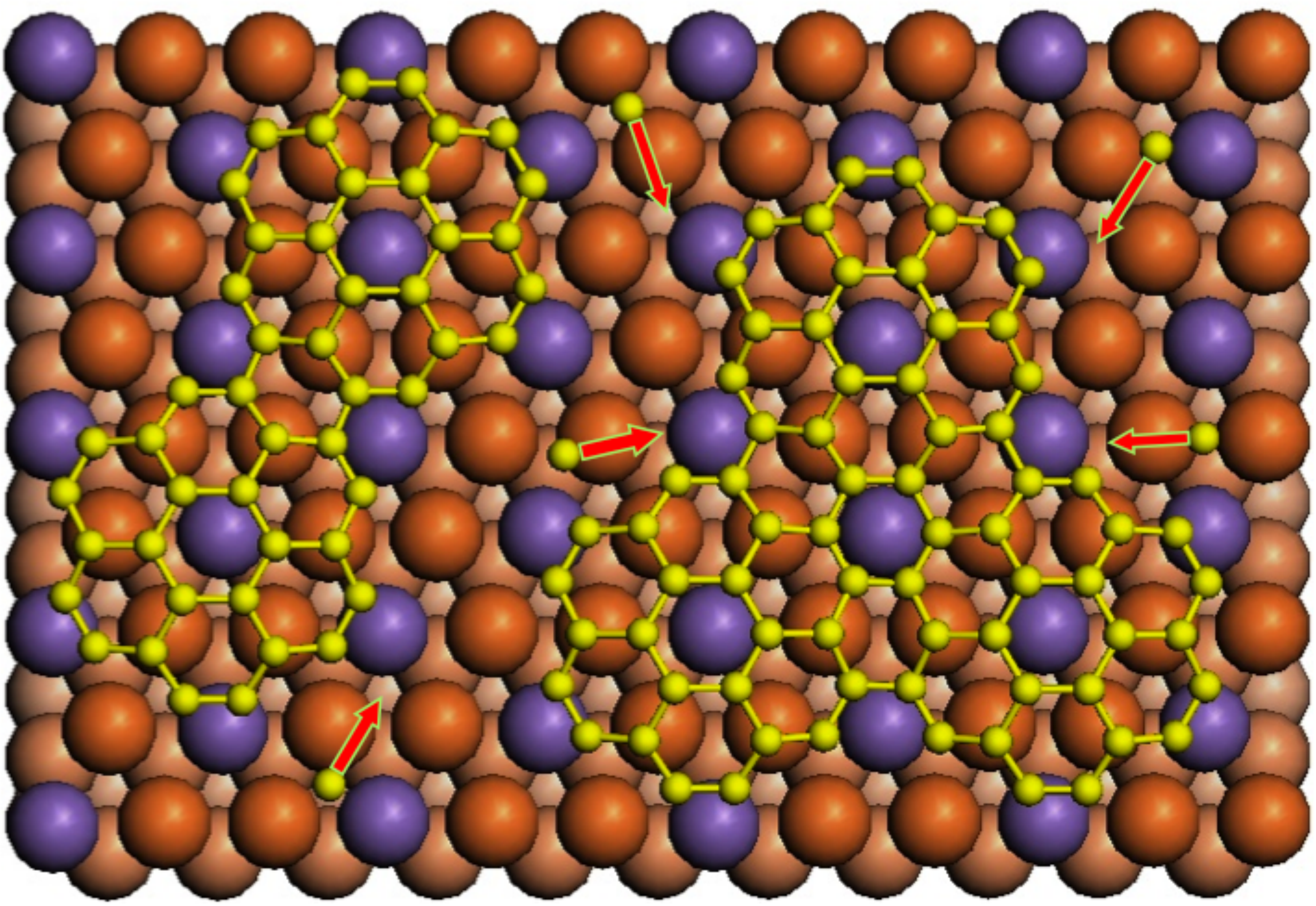}
 \end{center}
 \caption{(color online). Enlargement of coronene-seeded carbon islands via conventional CVD growth. The individual C adatoms supplied in the second step of the ``seed and grow'' kinetic pathway diffuse and attach to the nearby islands to fill the opening spaces, resulting in a larger graphene sheet with no GBs.}
 \label{figure4}
\end{figure}

It is important to note that, when a 7CR seed grows larger, the carbon clusters will still be at or close to the HSO, because an edge C atom prefers to reside at the bridge site between a Mn and a Cu atom. Because all the islands have nearly the same orientation and dome-like geometry, they will be able to make minimal local adjustments when they meet, and coalesce to form a single larger graphene sheet without GBs. Because of the lattice mismatch and its stronger interaction with the Mn-Cu(111) surface than with pure Cu(111), graphene may have a corrugated geometry, similar to that on Ru~\cite{martoccia_prl_2008}. Finally, the playground of using patterned substrates is not necessarily limited to the Mn-Cu(111) surface; other superstructured surface alloys with different transition metals beyond the ones already considered here are also worth exploring. The present study of graphene growth on patterned substrates via a two-step kinetic process thus opens the door towards a new and viable approach for mass production of single crystalline monolayer graphene.

This work was supported in part by the U.S. National Science Foundation (Grant No. 0906025 and No. 0854870), National Natural Science Foundation of China (Grant No. 11034006), and the UT/ORNL Joint Institute for Advanced Materials. W.G.Z. was supported in part by the U.S. Department of Energy, Basic Energy Sciences, Materials Science and Engineering Division. The calculations were performed at National Energy Research Scientific Computing Center (NERSC) of the U.S. Department of Energy.

\bibliographystyle{unsrt}

\begin{thebibliography}{99}

\bibitem{geim_natmater_2007} A. K. Geim and K. S. Novoselov, Nat. Mater. \textbf{6}, 183 (2007).

\bibitem{geim_science_2009} A. K. Geim, Science \textbf{324}, 1530 (2009).

\bibitem{kim_nature_2009} K. S. Kim \emph{et al.}, Nature \textbf{457}, 706 (2009).

\bibitem{coraux_newjphys_2009} J. Coraux \emph{et al.}, New J. Phys. \textbf{11}, 023006 (2009).

\bibitem{sutter_natmater_2008} P. W. Sutter, J. Flege, and E. A. Sutter, Nat. Mater. \textbf{7}, 406 (2008).

\bibitem{sutter_prb_2009} P. Sutter, J. T. Sadowski, and E. Sutter, Phys. Rev. B \textbf{80}, 245411 (2009).

\bibitem{ago_acsnano_2010} H. Ago, Y. Ito, N. Mizuta, K. Yoshida, B. Hu, C. M. Orofeo, M. Tsuji, K. Ikeda, and S. Mizuno, ACS Nano \textbf{4}, 7407 (2010).

\bibitem{kwon_nanolett_2009} S. Kwon, C. V. Ciobanu, V. Petrova, V. B. Shenoy, J. Bare\~{n}o, V. Gambin, I. Petrov, and S. Kodambaka, Nano Lett. \textbf{9}, 3985 (2009).

\bibitem{roth_surfsci_2011} S. Roth, J. Osterwalder, and T. Greber, Surf. Sci. \textbf{605}, L17 (2011).

\bibitem{oznuluer_apl_2011} T. Oznuluer, E. Pince, E. O. Polat, O. Balci, O. Salihoglu, and C. Kocabas, Appl. Phys. Lett. \textbf{98}, 183101 (2011).

\bibitem{li_science_2009} X. Li \emph{et al.}, Science \textbf{324}, 1312 (2009).

\bibitem{li_nanolett_2009} X. Li, W. Cai, L. Colombo, and R. S. Ruoff, Nano Lett. \textbf{9}, 4268 (2009).

\bibitem{gao_nanolett_2010} L. Gao, J. R. Guest, and N. P. Guisinger, Nano Lett. \textbf{10}, 3512 (2010).

\bibitem{sun_nature_2010} Z. Sun, Z. Yan, J. Yao, E. Beitler, Y. Zhu, and J. M. Tour, Nature \textbf{468}, 549 (2010).

\bibitem{li_acsnano_2011} Z. Li, P. Wu, C. Wang, X. Fan, W. Zhang, X. Zhai, C. Zeng, Z. Li, J. Yang, and J. Hou, ACS Nano \textbf{5}, 3385 (2011).

\bibitem{chen_prl_2010} H. Chen, W. G. Zhu, and Z. Y. Zhang, Phys. Rev. Lett. \textbf{104}, 186101 (2010).

\bibitem{li_jacs_2011} X. Li, C. W. Magnuson, A. Venugopal, R. M. Tromp, J. B. Hannon, E. M. Vogel, L. Colombo, and R. S. Ruoff, J. Am. Chem. Soc. \textbf{133}, 2816 (2011).

\bibitem{li_nanolett_transfer_2009} X. Li, Y. Zhu, W. Cai, M. Borysiak, B. Han, D. Chen, R. D. Piner, L. Colombo, and R. S. Ruoff, Nano Lett. \textbf{9}, 4359 (2009).

\bibitem{yu_natmater_2011} Q. Yu \emph{et al.}, Nat. Mater. \textbf{10}, 443 (2011).

\bibitem{ajayan_natmater_2011} P. M. Ajayan and B. I. Yakobson, Nat. Mater. \textbf{10}, 415 (2011).

\bibitem{yazyev_prb_2010} O. V. Yazyev and S. G. Louie, Phys. Rev. B \textbf{81}, 195420 (2010).

\bibitem{yazyev_natmater_2010} O. V. Yazyev and S. G. Louie, Nat. Mater. \textbf{9}, 806 (2010).

\bibitem{grantab_science_2010} R. Grantab, V. B. Shenoy, and R. S. Ruoff, Science \textbf{330}, 946 (2010).

\bibitem{huang_nature_2011} P. Y. Huang \emph{et al.}, Nature \textbf{469}, 389 (2011).

\bibitem{kim_acsnano_2011} K. Kim, Z. Lee, W. Regan, C. Kisielowski, M. F. Crommie, and A. Zettl, ACS Nano \textbf{5}, 2142 (2011).

\bibitem{an_acsnano_2011} J. An \emph{et al.}, ACS Nano \textbf{5}, 2433 (2011).

\bibitem{li_nanolett_2010} X. Li \emph{et al.}, Nano Lett. \textbf{10}, 4328 (2010).

\bibitem{rosenfeld_prl_1993} G. Rosenfeld, R. Servaty, C. Teichert, B. Poelsema, and G. Comsa, Phys. Rev. Lett. \textbf{71}, 895 (1993).

\bibitem{fetzer_2000} \emph{The Chemistry and Analysis of the Large Polycyclic Aromatic Hydrocarbons}, edited by  J. C. Fetzer (Wiley, New York, 2000).

\bibitem{schneider_ass_1999} J. Schneider, A. Rosenhahn, and K. Wandelt, Appl. Surf. Sci \textbf{142}, 68 (1999).

\bibitem{bihlmayer_prb_2000} G. Bihlmayer, Ph. Kurz, and S. Bl\"{u}gel, Phys. Rev. B \textbf{62}, 4726 (2000).

\bibitem{kresse_prb_1996} G. Kresse and J. Furthm\"{u}ller, Phys. Rev. B \textbf{54}, 11169 (1996).

\bibitem{blochl_prb_1994} P. E. Bl\"{o}chl, Phys. Rev. B \textbf{50}, 17953 (1994).

\bibitem{kresse_prb_1999} G. Kresse and D. Joubert, Phys. Rev. B \textbf{59}, 1758 (1999).

\bibitem{perdew_prl_1996} J. P. Perdew, K. Burke, and M. Ernzerhof, Phys. Rev. Lett. \textbf{77}, 3865 (1996).

\bibitem{methfessel_prb_1989} M. Methfessel and A. Paxton, Phys. Rev. B \textbf{40}, 3616 (1989).

\bibitem{lacovig_prl_2009} P. Lacovig, M. Pozzo, D. Alf\`{e}, P. Vilmercati, A. Baraldi, and S. Lizzit, Phys. Rev. Lett. \textbf{103}, 166101 (2009).

\bibitem{chen_prepare} W. Chen \emph{et al.}, in preparation.

\bibitem{chen_ssp_2008} J. G. Chen, C. A. Menning, and M. B. Zellner, Surf. Sci. Rep. \textbf{63}, 201 (2008).

\bibitem{barckholtz_jacs_1999} C. Barckholtz, T. A. Barckholtz, and C. M. Hadad, J. Am. Chem. Soc. \textbf{121}, 491 (1999).

\bibitem{martoccia_prl_2008} D. Martoccia \emph{et al.}, Phys. Rev. Lett. \textbf{101}, 126102 (2008).

\end{thebibliography}

\end{document}